\PassOptionsToPackage{
	breaklinks=true,
	colorlinks=true,
	citecolor=	blue,
	linkcolor=	blue,
	urlcolor=	blue}{hyperref}

\documentclass[letterpaper,aps,pra,twocolumn,english,showpacs,superscriptaddress]{revtex4-1}
\pdfoutput=1
\usepackage{hyperref} 
\usepackage{amsmath}
\usepackage{amsfonts}
\usepackage{footnote}
\usepackage{todonotes}
\usepackage[caption=false]{subfig}
\usepackage{xcolor}
\usepackage{graphicx}
\usepackage[squaren]{SIunits}
\makeatletter
\def\tagform@#1{\maketag@@@{\ignorespaces#1\unskip\@@italiccorr}}%
\let\orgtheequation\theequation%
\def\theequation{(\orgtheequation)}%
\makeatother

\newcommand{\efield}{\mathbf{\mathcal{E}}}

\begin{document}
\title{Two-Atom Rydberg Blockade using Direct $\boldsymbol{6S}$ to $\boldsymbol{nP}$ Excitation}

\author{A. M. Hankin}
%\email{ahankin@sandia.gov}
\affiliation{Sandia National Laboratories, Albuquerque, New Mexico 87185, USA}
\affiliation{Center for Quantum Information and Control (CQuIC), Department of
Physics and Astronomy, University of New Mexico 87131, USA}
\author{Y.-Y. Jau}
\author{L. P. Parazzoli} 
\author{C. W. Chou}
\author{D. J. Armstrong}
\affiliation{Sandia National Laboratories, Albuquerque, New Mexico 87185, USA}
\author{A. J. Landahl}
\author{G. W. Biedermann}
\email{gbieder@sandia.gov}
\affiliation{Sandia National Laboratories, Albuquerque, New Mexico 87185, USA}
\affiliation{Center for Quantum Information and Control (CQuIC), Department of
Physics and Astronomy, University of New Mexico 87131, USA}

\begin{abstract}
We explore a single-photon approach to Rydberg state excitation and Rydberg
blockade. Using detailed theoretical models, we show the feasibility of direct
excitation, predict the effect of background electric fields, and calculate the
required interatomic distance to observe Rydberg blockade. We then measure and
control the electric field environment to enable coherent control of Rydberg
states. With this coherent control, we demonstrate Rydberg blockade of two atoms
separated by 6.6(3)~\micro\meter. When compared with the more common two-photon
excitation method, this single-photon approach is advantageous because it
eliminates channels for decoherence through photon scattering and ac Stark
shifts from the intermediate state while moderately increasing Doppler
sensitivity.
\end{abstract}

\date{\today}
\maketitle
\section{Introduction}

The large polarizability of high principal quantum number $n$ Rydberg states
gives rise to exotic many-body interactions as well as an extreme sensitivity to
the electric field environment. Precision spectroscopy of such states allows for
a variety of exciting demonstrations in metrology, fundamental quantum
mechanics, and quantum information. For example, cold Rydberg atoms employed as
near-surface electric field sensors enable characterization of both field
amplitude and source. This includes experiments that explore near-surface field
spectral density \cite{Martin2013_PRA}, induced dipole moments for surface
adatoms \cite{Spreeuw2010PRA}, and insulator charging on an atom chip
\cite{Martin2012_PRA}. Large Rydberg state polarizabilities also enable
long-range electric dipole-dipole interactions (EDDIs) between Rydberg atoms,
yielding strongly correlated systems through the Rydberg blockade effect.
Recent experiments use Rydberg blockade to observe entanglement between neutral
atoms \cite{Grangier2009_NaturePhys,Browaeys2010_PRL}, a controlled-NOT quantum
gate \cite{Saffman2010PRL}, and collective many-body Rabi oscillations
\cite{Kuzmich2012_NatPhys}. These advances parallel an ever-evolving approach to
Rydberg state control. In this paper, we demonstrate Rydberg blockade using a
unique single-photon excitation approach to precision Rydberg spectroscopy.

The ionization threshold for ground-state alkali-metal atoms ranges from 3.9 to
5.4~eV, setting the energy scale for excitation to high-lying Rydberg states.
In practice, this is commonly accomplished with two-photon excitation, where the
ground and Rydberg states couple together through an intermediate state
\cite{SaffmanWalker2010}. Two-photon excitation avoids deep, ultraviolet (UV)
wavelengths making the implementation technologically simpler. However, photon
scattering and ac Stark shifts from the intermediate state introduce avenues for
decoherence, frequency noise, and dipole forces complicating two-photon
experiments \cite{SaffmanWalker2012__AMOPhys}.  Minimizing photon scattering is
of particular importance when using Rydberg-dressed atoms to create tunable,
long-lived, many-body interactions in a quantum gas \cite{Rolston2010PRA}.  For
example, the adiabatic quantum optimization protocol described in
\cite{Keating2013PRA}, is predicted to achieve a substantially higher fidelity
in the absence of photon scattering from the intermediate state. At present,
studies of Rydberg blockade with a single-photon transition are rare.  Previous
work utilized single-photon excitation with pulsed UV lasers to perform
high-resolution spectroscopy of Rydberg states \cite{Deiglmayr2013_PRA} and to
detect Rydberg blockade as a bulk effect \cite{Gould2004_PRL}.  Still, direct
excitation using a continuous-wave (cw) UV laser, which would allow for coherent
control of single atoms, has not been demonstrated.

Here, we show coherent control of blockaded 84$P_{3/2}$ states of two single
$^{133}$Cs atoms using a cw UV laser at 319~nm.  Construction of this UV laser
is informed by calculations for the required wavelength and intensity.  With
over 300~mW of 319-nm light at the output of the laser, we demonstrate
a Rabi frequency of over 2~MHz with this approach, in agreement with our
predictions for resonance frequency and oscillator strength. Given this success,
we further develop our model to determine a regime for observing Rydberg
blockade between two atoms, and we observe and analyze Rydberg blockade for the
84$P_{3/2}$ state. 

This paper is organized in the following way: In
\autoref{sec:directExciteModel}, we establish a detailed model for single-photon
excitation to Rydberg P states that includes predictions for the Rydberg
spectrum and oscillator strengths. We next use this model to design a cw UV
laser system, the details of which are found in \autoref{sec:rydbergLaser}. In
\autoref{sec:experiment}, we describe the experimental technique used to trap
and control two atoms in close proximity.  In \autoref{sec:efield}, we use our
single-atom control in combination with the UV laser system to measure the
background electric field inherent to our apparatus. We then implement active
control and suppression of the electric field to enable coherent control of the
atom. In \autoref{sec:rydbergBlockade}, we present our study of the Rydberg
blockade, including a model of the Rydberg spectrum for two atoms as a function
of interatomic spacing and an experimental demonstration of the blockade effect
in this system. We conclude with applications where the single-photon excitation
approach is expected to excel.

\section{Single-photon excitation model}
\label{sec:directExciteModel}
\begin{figure}%[t]
		\includegraphics{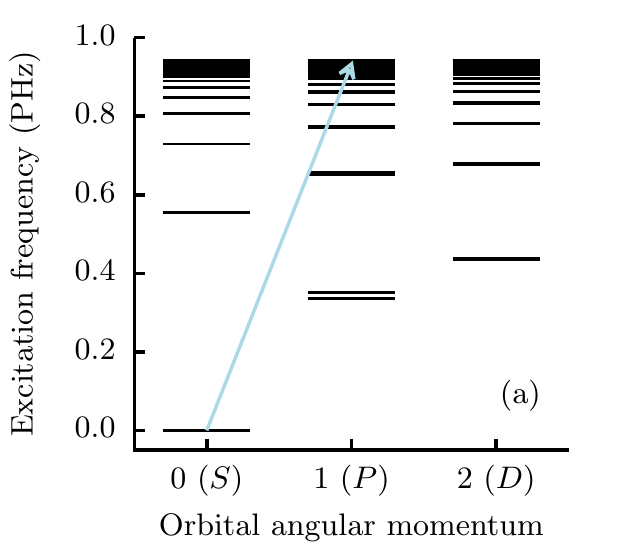}

		\includegraphics{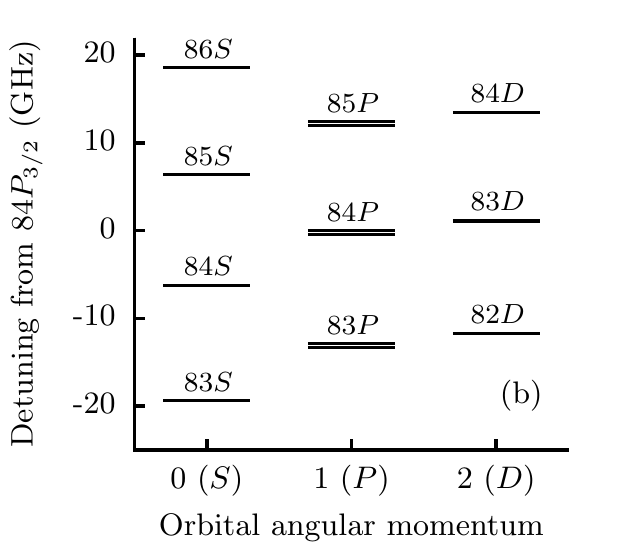}
	\caption{ (Color online)
	Spectrum of cesium using QDT.  (a) Spectrum from the
	ground state through $n=100$. The experimental work in this paper focuses on
	the $6S_{1/2} \rightarrow$ $84P_{3/2}$ transition, labeled by the blue
	arrow.	(b) Detailed spectrum near $84P_{3/2}$. Fine structure splitting in
	the $nP_{j}$ and $nD_j$ states is included. $n$P states appear broader
	due to splitting between $nP_{1/2}$ and $nP_{3/2}$. Fine structure
	between the $nD$ states is $\sim100$~MHz and therefore not well
	resolved on this scale.
	\label{fig:energyLvlDiagram}
	}
\end{figure}

We use a theoretical model for single-photon excitation of high-$n$ states to
accurately calculate the Rydberg spectrum and 6$S_{1/2}\rightarrow nP$
transition oscillator strengths. The spectra of alkali-metal atoms are predicted
with high precision by quantum defect theory (QDT) \cite{Seaton1983}. Using QDT,
the energies of bound electronic states $E$ are given by
\begin{equation}
	E(n,\ell,j) = E_{\infty} - \frac{R_{\text{Cs}}}{(n - \delta(n,\ell,j))^2},
	\label{eq:quantumdefect}
\end{equation}
where $\ell$ is the orbital angular quantum number, $j$ is the total angular
momentum quantum number, $E_{\infty}$ is the ionization threshold energy,
$R_{\text{Cs}}$ is the Rydberg constant for cesium, and $\delta$ is the quantum
defect. A method for calculating $\delta(n,\ell,j)$ is found in \cite{Weber1986}
where the observed Rydberg spectra are fit to a power series in $n$.  The
spectrum of $^{133}$Cs calculated with \autoref{eq:quantumdefect} is found in
\autoref{fig:energyLvlDiagram}. The optical frequency for excitation directly
from $6S_{1/2}$ to $84P_{3/2}$ is calculated to be $941\, 030$~GHz.

Given the transition frequency, we require the transition oscillator strength
$f$ to determine if single-photon excitation is feasible with current
technology.  A semiempirical method for calculating $f$ is found in
\cite{Fabry1976}. Computing $f$ requires knowledge of the radial wave functions
and the associated radial matrix elements $\langle n'\ell'j'|r|n\ell j\rangle$.
The radial wave function is calculated by substituting the energies predicted by
QDT into Schr\"{o}dinger's equation and $\langle n'\ell' j|r|n\ell j\rangle$ is
then computed through numerical integration.  We find the oscillator strengths
for the $6S_{1/2}$ to $84P$ transitions are $f(6S_{1/2}\rightarrow
84P_{3/2}) = 6\times10^{-8}$ and $f(6S_{1/2}\rightarrow 84P_{1/2}) =
2\times10^{-12}$. The $10^4$ order of magnitude difference between the
calculated oscillator strengths is a property that is unique to cesium when
compared with the other alkali-metal atoms.  The divergence of the principal-series
doublet (6$S_{1/2}\rightarrow nP_{3/2,1/2}$) oscillator strength ratio for large
$n$ is a well-known phenomenon that arises with the inclusion of spin-orbit
effects and the core polarizability
\cite{Fermi1930,Fabry1976,Stroke1981,Weisheit1972_PRA}.  This result favors
exciting to $nP_{3/2}$ states in the interest of reducing laser power
requirements.

While the oscillator strength determines the scaling of Rabi frequency~$\Omega$
with laser intensity, it does not directly set a lower limit on laser power.
Instead, we must consider limits on the Rydberg laser waist and the coherence
time between the Rydberg state $|r\rangle$  and the ground state $|g\rangle$ set
by experimental conditions. For a reasonable atom temperature of
\unit{10}{\micro\kelvin} and a trap waist of \unit{1}{\micro\meter}, the atom
velocity spread limits the linewidth to order \unit{100}{\kilo\hertz} and the
spatial spread to order \unit{1}{\micro\meter}. We target a Rabi frequency  on
the order of $\Omega/2\pi = 1$~MHz and a laser waist of \unit{10}{\micro\meter}
to avoid decoherence and intensity fluctuations due to atom motion. Combining
the limitation on $\Omega$ with the targeted waist, we find that 16~mW of 319~nm
light is sufficient to observe state evolution that is dominated by coherent
dynamics. The design for the cw UV laser described in the following section
surpasses this requirement.

\section{Rydberg Laser}
\label{sec:rydbergLaser}

\begin{figure*}
	\includegraphics{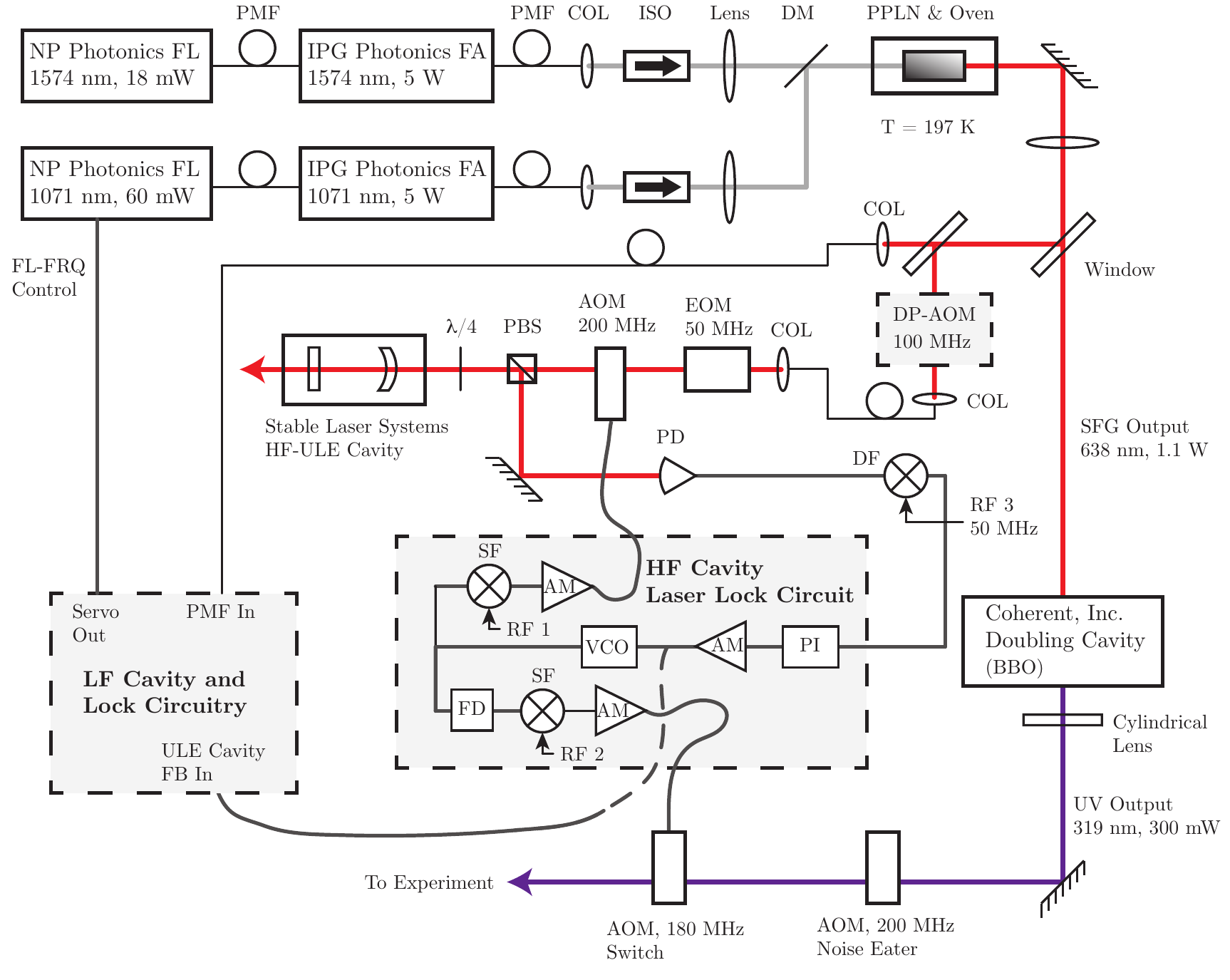}
	\caption{ (Color online)
	Diagram of UV laser system. Frequency summing in a PPLN crystal of two fiber
	laser systems yields 638~nm light and frequency doubling of this light in
	BBO generates UV light at 319~nm. Laser frequency is stabilized using a
	multistaged servo with an ultra-low expansion (ULE), high finesse (HF)
	cavity as the primary reference. 
	$\lambda/4$---quarter-wave plate,
	AM---voltage amplifier,
	COL---fiber collimation package,
	DF---difference frequency,
	DM---dichroic mirror,
	DP-AOM---double-pass acousto-optic modulator system,
	EOM---electro-optic modulator,
	FA---fiber amplifier,
	FB---feedback,
	FD---frequency doubling,
	FL---fiber laser, 
	FL-FRQ---voltage control of fiber laser frequency,
	ISO---optical isolator,
	LF---low finesse,
	PD---photodiode,
	PI---proportional-integral feedback,
	PMF---polarization maintaining fiber,
	SF---sum frequency,
	RF---radio frequency source,
	VCO---voltage controlled oscillator.
	}
	\label{fig:uvLaserDiagram}
\end{figure*}

The cw UV laser is constructed using sum frequency generation (SFG) followed by
frequency doubling. A similar approach tailored for 313~nm is found in
\cite{Wilson2011APB}. We first produce 638~nm light using SFG and then generate
the 319-nm light via frequency doubling.  The SFG begins with 1574- and
1071-nm fiber laser sources with 18- and 60-mW output powers, respectively.
Both lasers seed commercial 5-W fiber amplifiers and the resulting light is
combined and passed through a periodically poled lithium niobate (PPLN) crystal
generating 1.1~W of 638-nm light.  The output of the PPLN crystal is frequency
doubled from 638 to 319~nm with a BBO ($\beta$-BaB$_2$O$_2$) crystal and
results in greater than 300~mW at this wavelength. From the spectrum shown in
\autoref{fig:energyLvlDiagram}, we predict the laser's frequency can be tuned to
reach 84$P$ through 120$P$. Upon exiting the doubling cavity, the beam is shaped
into a Gaussian profile, passed through two AOMs for intensity stabilization and
switching, and focused down on the two atoms with a measured 1/e$^2$ radius of
\unit{12.9(3)}{\micro\meter}.  This results in the two atoms experiencing a
maximum intensity of $60$~kW/cm$^2$ after accounting for the losses incurred at
each optic. 

The frequency of the Rydberg laser is stabilized to an ultra-low expansion, high
finesse (HF) cavity \footnote{The cavity has a finesse of $20\,000$ at $638$~nm.
The cavity temperature is stabilized and kept under vacuum at $10^{-7}$~ Torr to
reduce environmental effects driving the cavity resonance frequency.} at 638~nm
via a multistage servo architecture (\autoref{fig:uvLaserDiagram}). A direct
lock to the cavity frequency reference is precluded by characteristics of the
fiber laser sources. Frequency noise on the 638~nm light exceeds the 75~kHz
cavity linewidth as well as the bandwidth of the frequency control of the fiber
lasers. We overcome this by dividing low and high bandwidth frequency
stabilization into two paths.

\begin{figure*}%[htbp]
	\subfloat[top view]{
		\includegraphics{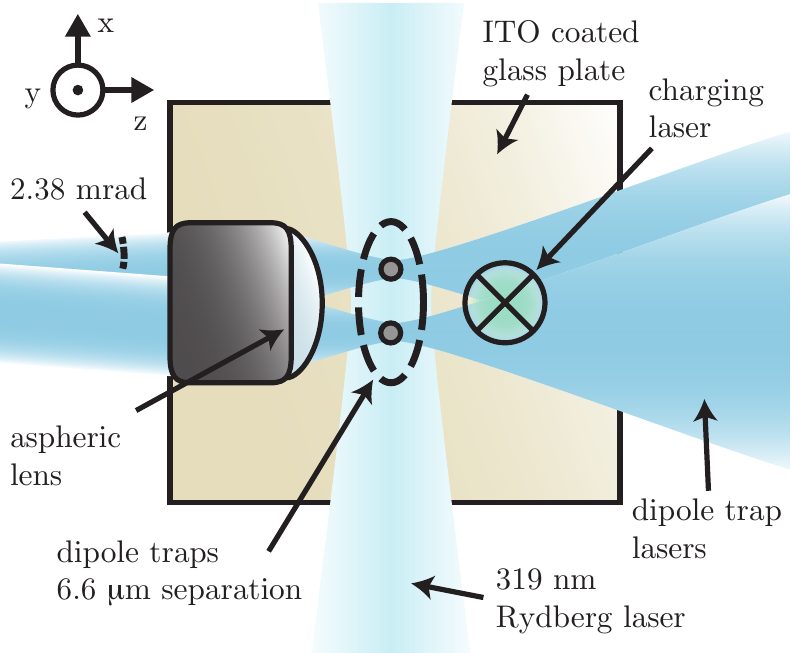}}
	\hspace{1cm}
	\subfloat[side view]{
		\includegraphics{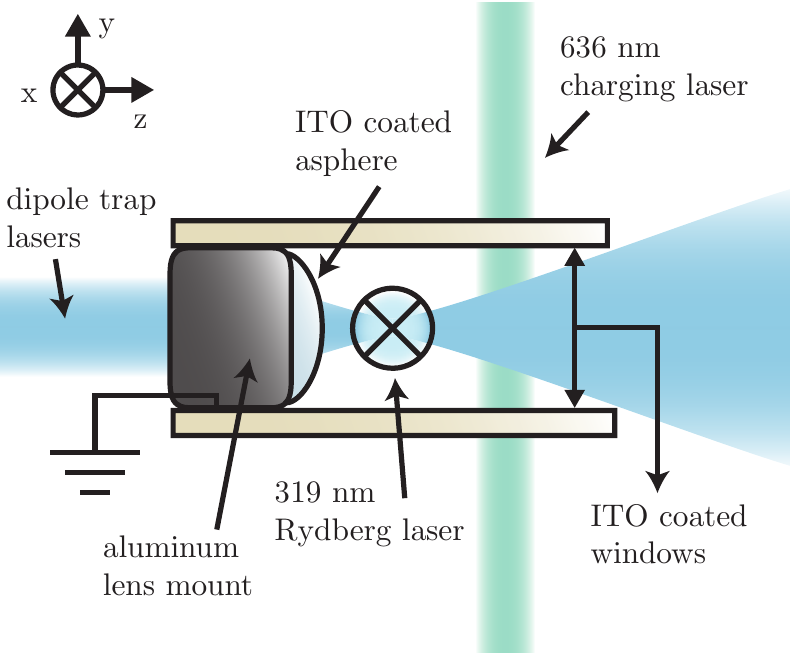}}
	\caption{ (Color online)
	Diagram of the atom trapping region. Two collimated 938-nm dipole trap beams
	with a 2.38-mrad relative angle pass through an aspheric lens resulting in
	two traps separated by 6.6(3)~$\mu$m at the focal plane.  A
	319-nm laser, used to excite to Rydberg states, is focused down to a
	\unit{12.9(4)}{\micro\meter} waist at the location of the atoms. The
	aspheric lens has a 112-nm ITO coating on the side facing the dipole traps
	and an anti-reflection (AR) coating for 852~nm on the opposite side. An
	aluminum cylinder is fixed concentric to the AR-coated side to shield
	against charging of this dielectric.  The resulting assembly is fixed
	between two ITO-coated glass plates with a vacuum compatible, conductive
	epoxy. Each plate has an ITO coating on the side closest to the traps. The
	entire assembly is grounded.  The top ITO plate is not shown in (a) for
	clarity. A 636-nm charging laser beam generates controlled charging on the
	ITO plates, which grants leverage over the background electric field
	environment. Figures are not to scale.
	} 
	\label{fig:experimentSetup}
\end{figure*}

Low-bandwidth frequency control is implemented by first stabilizing the 638-nm
light to a low finesse (LF) cavity with a 5~MHz linewidth. This frequency
stabilization stage narrows the laser linewidth through feedback to the 1071~nm
fiber laser.  Next, we split a small fraction of the 638-nm light along an
optical path used to monitor and stabilize the light's frequency to the HF
cavity. Because the LF cavity lock has narrowed the 638-nm linewidth, we are
able to directly lock the light along this path to the HF cavity. The HF cavity
lock system consists of an electro-optic modulator (EOM), an AOM, and the HF
cavity. The locking error signal is generated using the Pound-Drever-Hall
technique where the EOM modulates the phase of the light while the response is
monitored in reflected cavity signal. We then feedback to the drive frequency of
the AOM to stabilize the frequency of the light to the cavity. However, because
this AOM is not placed in the primary, high-power 638-nm beam path, the laser
frequency at the atom does not directly benefit from the servo.  This leaves the
primary laser beam path susceptible to drifts in the LF cavity length with
changes in temperature, pressure, and humidity. We avoid this issue using a
low-bandwidth feedback loop that adjusts the LF cavity length to stabilize the
frequency of the 638-nm light to the HF cavity resonance. The result of this
portion of the locking system is a 638-nm linewidth of no more than 200~kHz
along the high-power beam path.

The linewidth of the 319-nm light is narrowed further using high-bandwidth
feed-forward control. While it is possible to use a closed-loop servo to
directly stabilize the laser frequency by placing the control AOM directly in
the high-power 638-nm beam path, we choose a feed-forward approach in favor of
maximizing the UV power. The feed-forward control is accomplished by splitting
the radio frequency (RF) signal used to stabilize the 638-nm laser to the HF
cavity along a second path. This new path applies high-bandwidth corrections to
the 319-nm laser frequency by modulating the drive frequency of an AOM in the UV
beam path. The circuit compensates for the frequency doubling that occurs in the
optical domain at the BBO crystal by placing a frequency doubler in the second
RF path. This feed-forward architecture transfers the high-frequency content of
the HF cavity lock to the UV light.

\section{Experimental Apparatus}
\label{sec:experiment}

We create two optical dipole traps for quantum control of single cesium atoms in
an all glass, ultra high vacuum cell.  The traps are generated inside a partial
Faraday cage to control stray electric fields that perturb the Rydberg states.
Rydberg excitation of these atoms is detected via atom loss as Rydberg states
are ejected from the trap. By carefully tuning the optical parameters to
minimize heating, we are able to reuse the same atoms for multiple experiments.
The experimental system used to trap and probe two single $^{133}$Cs atoms was
built by modifying the setup described in \cite{Parazzoli2012PRL}. The atoms are
confined with a well defined separation, in two far off-resonant dipole traps
using 938~nm light.  The trapping light passes first through an acousto-optical
modulator (AOM) and then through an in-vacuum aspheric lens with a 2.76~mm focal
length. By driving the AOM at two frequencies (74.6 and 85.4 MHz) we generate
two 8~mW beams whose propagation directions deviate by 2.38~mrad.  This AOM-lens
system results in two dipole traps separated by \unit{6.6(3)}{\micro\meter}
\footnote{The quoted error is derived from focal length and atom motion
uncertainty.}. Both traps have a \unit{1.26(1)}{\micro\meter} waist and a
21.1(1)~MHz trap depth for the atomic ground state. Once trapped, the atoms have
a vacuum limited trap lifetime of approximately 7~s. We source the 938~nm light
from a distributed feedback laser diode and find it necessary to filter elements
of 852 and 895~nm from this laser ($D_2$ and $D_1$ transitions in $^{133}$Cs)
to avoid excessive heating that inhibits stable trapping.

The atoms are trapped 2.16~mm from the lens surface where background electric
fields can be problematic for coherent control of Rydberg atoms
\cite{Browaeys2013__PRL}. We suppress these fields by coating the surface of the
lens closest to the atom with an 112~nm layer of indium tin oxide (ITO). This
transparent yet conductive coating is grounded to dissipate charging.  To
further protect against the influence of external electric fields, we surround
the trapping region with a partial Faraday cage in vacuum by mounting the lens
between two parallel glass plates that are also coated with ITO
(\autoref{fig:experimentSetup}). Using finite element analysis to approximate
the solution to Laplace's equation for the electric potential, we calculate that
this geometry suppresses electric fields external to the system by a factor of
1000. 

The atoms are loaded into the dipole traps from a magneto-optical trap (MOT).
The dissipative scattering force generated by the MOT cools atoms into the
conservative pseudo-potential of these traps. Once captured, the atoms continue
to fluoresce on the $D_{2}$ transition of $^{133}$Cs (6$S_{1/2}\rightarrow$
6$P_{3/2}$), and we spatially discriminate this signal to detect a loading
event. This light is collected by the same aspheric lens used to produce the
dipole traps and a dichroic mirror separates the 938~nm trapping light from this
852~nm fluorescence of the $D_2$ transition. After reflecting off the dichroic,
the two beams of fluorescence are imaged at a plane coincident with a gold knife
edge.  The knife edge is positioned such that the image from one atom is
reflected off of the gold surface, while the image from the other passes. Next,
the fluorescence of each atom is coupled into separate \unit{9}{\micro\meter}
core fibers \footnote{Fibers are single mode at 1550~nm.} that feed separate
avalanche photodiodes (APDs).  We find that this core size yields a near-optimal
signal-to-noise ratio for single-atom detection in our apparatus.  The technique
we use to split the fluorescence beams is similar to the one developed by
\cite{Browaeys2010_PRL}.  We adjust the MOT cloud density to operate both traps
in the collisional blockade regime such that loading is limited to a maximum of
one atom~\cite{Grangier2002PRL}. By waiting for a coincidence of bright
fluorescence signals from both APDs, we load single atoms simultaneously in both
traps. Once loaded, we switch off the loading process by extinguishing the MOT
lasers and the quadrupole magnetic field. After a 10-ms wait period that allows
for the magnetic field environment to stabilize and the MOT cloud to dissipate,
we prepare the same atoms for the single-photon excitation experiment.

\begin{figure}%[htbp]
	\begin{center}
		\includegraphics{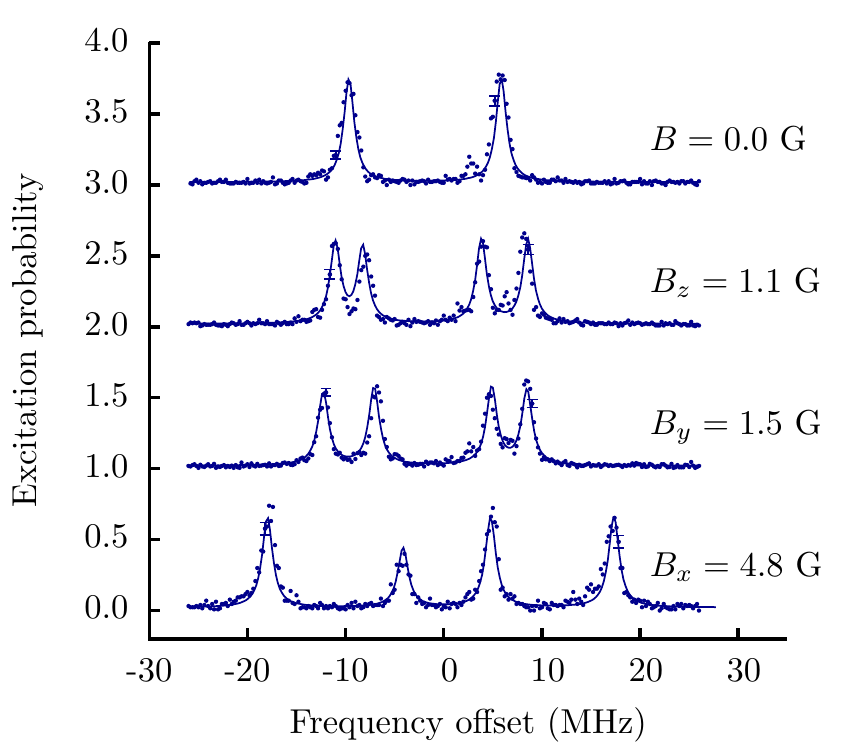}
	\end{center}
	\caption{ (Color online)
	Spectroscopy of the 84P$_{3/2}$ state for various bias magnetic fields
	$\mathbf{B}$. From top to bottom the spectrum is shown for $\mathbf{B}$ =
	(0,0,0), (0,0,$B_z$), (0,$B_y$,0), and ($B_x$,0,0) where the $x$, $y$, and
	$z$ axes are labeled in \autoref{fig:experimentSetup}.  The solid blue line
	represents a model for the spectrum that includes perturbations to the state
	due to the presence of magnetic and electric fields. A fit of the model to
	the data using the magnitude and direction of $\mathbf{\mathcal{E}}$ as free
	parameters indicates the presence of a 6.35(5)~V/m electric field collinear
	with the $z$ axis (normal to the dipole trap lens) with a $\pm20\degree$
	uncertainty. The spectra are offset by multiples of 1.0 on the $y$ axis of
	the plot for clarity.
	}
	\label{fig:efieldmeasure}
\end{figure}

Before excitation, we further cool the atoms and prepare them in $|6S_{1/2}, F
= 4, m_F = 0\rangle$. The atoms are cooled to \unit{16.1(1)}{\micro\kelvin}
using polarization-gradient cooling \cite{Wineland1992_JOSAB}. The experimental
details for this cooling process are found in \cite{Parazzoli2012PRL}. For state
preparation into $|F=4, m_F=0\rangle$, a quantization axis is defined with a
4.8~G magnetic field and the atoms are illuminated with $\pi$-polarized light
resonant with $F=4\rightarrow F'=4$ on the $D_1$ transition ($6S_{1/2}
\rightarrow 6P_{1/2}$). We find the large bias magnetic field necessary to
obtain an optical pumping efficiency of 95(2)\% into the target state
\cite{Regal2012_PRX}. 

\begin{figure*} 
	\includegraphics{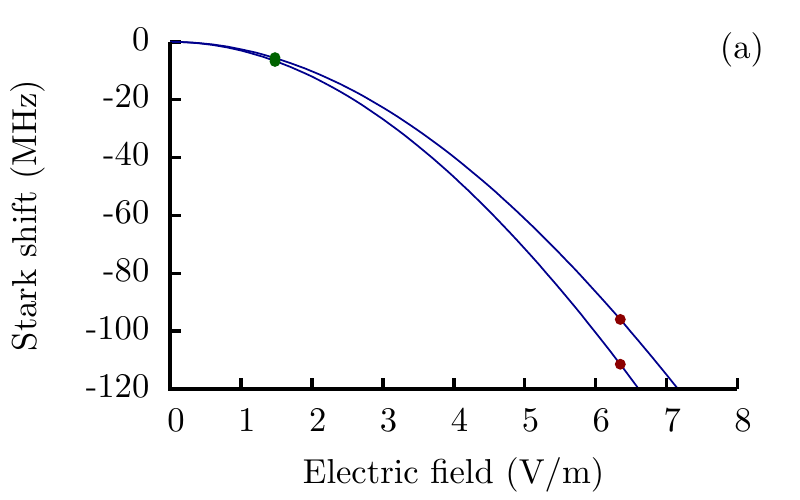}
	\hspace{.2in}
	\includegraphics{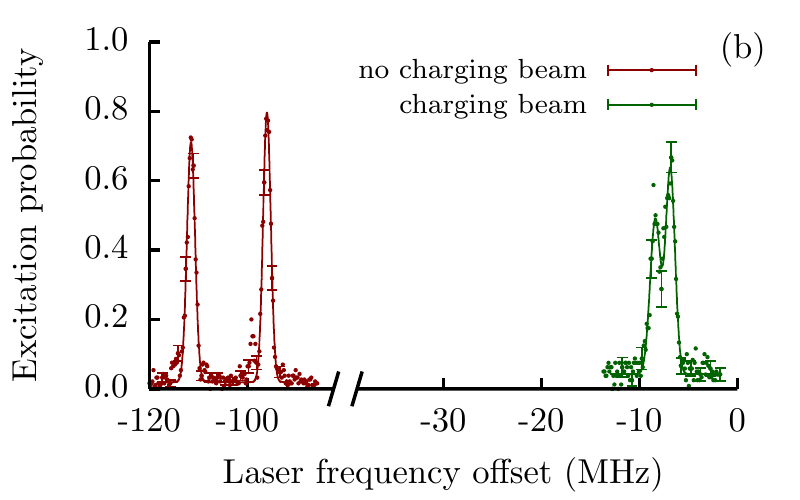}
	\caption{ (Color online)
	Electric field reduction via controlled surface charging with the charging
	laser. (a) Calculated Stark shift in the frequency of the 84$P_{3/2}$ state
	with no applied bias magnetic field. (b) Observed spectrum with (green
	points) and without (red points) the charging laser beam. Measured resonance
	frequencies are plotted directly on (a) using the same color scheme. The
	presence of the charging beam reduces the electric field magnitude to
	1.5(1)~V/m. Error bars shown are representative.
	}
	\label{fig:charging}
\end{figure*}

The cold, $|F=4,m_F=0\rangle$ atoms are the starting point for the direct
excitation experiments. These experiments begin 1~ms following state preparation
using the 319~nm laser described in \autoref{sec:rydbergLaser}.  Just before
excitation, the atoms are released into free-flight to avoid perturbations in
the ground-state energy from the dipole trap.  They are then recaptured
\unit{1}{\micro\second} after the excitation pulse extinguishes. The results of
these experiments are found in Secs.\ \ref{sec:efield} and
\ref{sec:rydbergBlockade}. We detect population in the Rydberg state by
taking advantage of the different trapping forces experienced by the
ground and Rydberg states for an atom in a red-detuned optical dipole trap;
while the ground state of the atom is trapped, a Rydberg state will experience
an anti-trapping potential \cite{SaffmanWalker2010} causing it to quickly eject
from the dipole trap and allowing atom loss to signal Rydberg excitation
\cite{Grangier2009_NaturePhys}. 

We check for atom loss 1~ms after excitation, allowing ample time for a Rydberg
state to leave the trapping region. For this check, we probe for atom
fluorescence on the $D_{2}$ cycling transition ($F = 4 \rightarrow F' = 5$) for
\unit{500}{\micro\second} with two-counter propagating detection lasers that are
directed along the y-axis (\autoref{fig:experimentSetup}) superimposed with
repump light on the $F = 3 \rightarrow F' = 4$ transition.  A detection beam
detuning of 5.0~MHz and an intensity of 17~mW/cm$^2$ optimally trades heating
rate and signal-to-noise ratio. The APD that monitors fluorescence during
detection measures an average number of photon counts of 8.4 and 0.51 for the
bright (atom present) and dark state (no atom) respectively. With a
discriminator equating the bright-state with a measurement of greater than two
counts, we achieve single-shot atom presence detection with a 95\% fidelity. We
optimize the position of the gold knife edge to homogenize the response of the
two traps.  The measured mean bright-state counts differ by 5\% between the two
traps and the atom presence detection fidelity differs by 0.5\%.  While
increasing the detection pulse duration improves fidelity it also increases
heating due to photon scattering which leads to atom loss. Choosing a shorter
detection pulse reduces the probability to lose an atom allowing for the reuse
of an atom over multiple iterations.

The trap-loading phase of the experiment can limit our bandwidth by consuming
around 97\% of the duty cycle. To mitigate this reduced bandwidth we capitalize
on atom reuse by employing a field-programmable gate array (FPGA) based control
system for high-speed Boolean logic.  The FPGA control system allows us to
increase our data rate from the 1~Hz level to a maximum of 70~Hz for single-atom
experiments. The logic implements a high speed flow chart that responds
appropriately to outcomes of the detection sequence.  When the atom is not
detected during the check sequence, a more robust check for atom presence is
performed immediately. The robust check stage consists of a maximum of three MOT
beam pulses identical to the pulse used for detection. If the atoms are observed
during the detection pulse or any of the pulses during the check sequence, then
the entire experimental cycle is repeated, skipping the rate-limiting
trap-loading phase. Otherwise, the MOT is repopulated to reload the traps.
Heating induced by the detection beam is minimized by using the minimum number
of pulses required to verify atom presence.  The actual data rate depends on the
specifics of the experiment. We achieve a 70-Hz data rate when performing
single-atom, state-selective, lossless detection experiments similar to those
described in \cite{Browaeys2011_PRL,Chapman2011__PRL}. When implementing
experiments with loss-based detection techniques, the data rate is lower and
depends on the experiment specific loss rate. For example, the average data rate
for the experiments detailed in Secs.\ \ref{sec:efield} and
\ref{sec:rydbergBlockade} are 2~Hz (\autoref{fig:efieldmeasure}), 0.7~Hz
(\autoref{fig:rydbergBlockade}, single atom), and 0.3~Hz
(\autoref{fig:rydbergBlockade}, two atoms). The data rate in these experiments
can, in principle, be greatly improved by combining lossless detection
techniques with the transfer of Rydberg state population to $|6S_{1/2}, F
= 3, m_F = 0\rangle$~\cite{Browaeys2010_PRL}.

\section{Electric field environment}
\label{sec:efield}

Rydberg electron wave functions scale to extremely large sizes with increasing
$n$. Consequently, dipole matrix elements between adjacent states grow as well,
scaling like $n^2a_0e$ \cite{gallagher2005rydberg}, where $a_0$ is the Bohr
radius and $e$ is the elementary charge. This in turn implies extreme
sensitivity to dc electric fields due to increasingly large electric
polarizabilities. We calculate that the 84$P_{3/2}$ state polarizability
$\alpha_r$ is on the order of $10^{11}$ times larger than that of the ground
state. We use this large polarizability to measure the electric field
environment at the dipole traps by studying the spectrum of the Rydberg state.

The Rydberg spectrum we measure for 84$P_{3/2}$ is shown in
\autoref{fig:efieldmeasure}.  The excitation experiment uses a 580-ns UV laser
pulse over a 50-MHz laser frequency scan range.  While we expect a single peak
in the absence of any external perturbations, the observed spectrum consists of
two nondegenerate peaks.  The observed degeneracy breaking for the zero
magnetic field condition ($B = 0$) in \autoref{fig:efieldmeasure} is
the result of a background electric field that shifts the resonances through the
dc Stark effect.  To further our understanding of the background electric field
source, we characterize the Rydberg spectrum at several different bias magnetic
field directions and compare the result with a detailed model.

We model the splitting in the four Rydberg resonances by including the
perturbing effects of electric and magnetic fields. The relative splitting is
calculated by diagonalizing the matrix of the total Hamiltonian,
\begin{equation}
	H = H_{\text{atom}} + H_{\text{Stark}} + H_{\text{Zeeman}}.
	\label{eq:atomicSpectrum}
\end{equation}
Here, $H_{\text{atom}}$ is the unperturbed Hamiltonian of a single atom and it's
matrix elements can be constructed using QDT \cite{Weber1986}.  The final two
terms are given by $H_{\text{Stark}} = -\mathbf{\boldsymbol\mu_{d}} \cdot
\mathbf{\mathcal E}$ and $H_{\text{Zeeman}} = -\mathbf{\boldsymbol\mu_{m}} \cdot
\mathbf{B}$. Here the electric and magnetic dipole moment operators are given by
$\mathbf{\boldsymbol\mu_{d}}$ and $\mathbf{\boldsymbol\mu_{m}}$ respectively,
$\mathbf{\mathcal{E}}$ is a dc electric field, and $\mathbf{B}$ is a dc magnetic
field. While the strength and direction of $\mathbf{B}$ is controlled using
three sets of Helmholtz coils, the electric field is a background intrinsic to
our system.  We diagonalize \autoref{eq:atomicSpectrum} for a set of states
large enough to ensure convergence of the eigenvalues and eigenvectors of $H$
over the chosen electric field range. This includes states where $n$ ranges from
81 to 89, and $\ell$ ranges from 0 to 6. The matrix elements for
$H_{\text{Stark}}$ and $H_{\text{Zeeman}}$ can be calculated with techniques
described in \cite{Zimmerman1979_PRA,Hirano1985_IOP}. A comparison of this
theoretical model with experimental data is shown in
\autoref{fig:efieldmeasure}.  Using the direction and strength of the electric
field as a free parameter, we find that there is a 6.35(5)~V/m electric field at
the location of the atom, pointed along a direction perpendicular to the lens
surface (\autoref{fig:experimentSetup}). This field direction indicates charging
of the dipole trap lens.

\begin{figure}%[htbp]
	\includegraphics{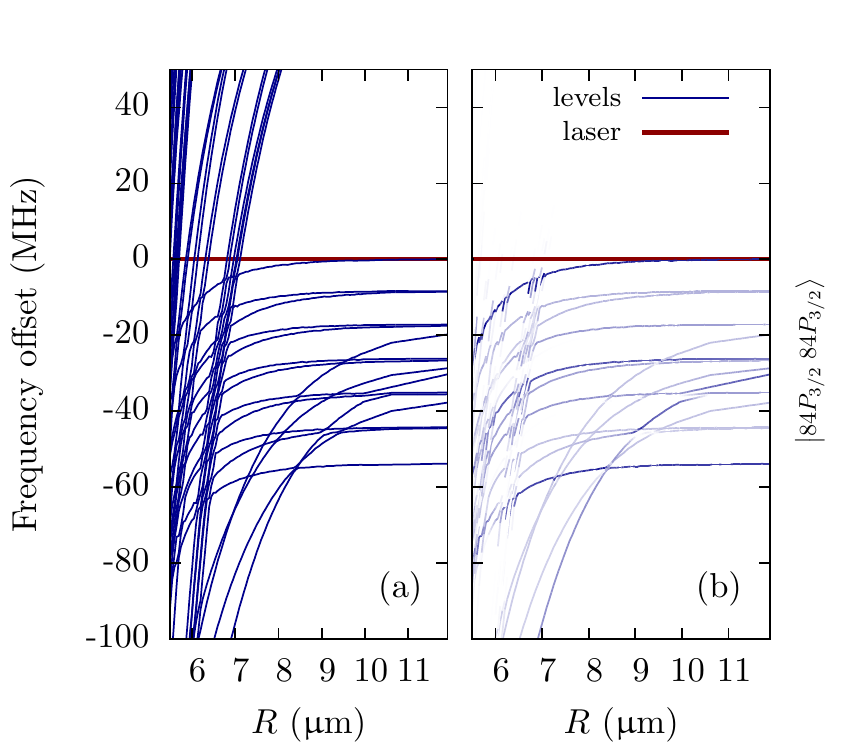}
	\caption{ (Color online)
	Numerical calculation for the $84P_{3/2}+84P_{3/2}$ spectrum
	versus interatomic separation $R$. The laser is resonant with the
	unperturbed ($R\rightarrow \infty$) state $|rr\rangle$, where $r \rightarrow
	\{84P_{3/2}, m_j=3/2\}$. (a) Frequency offset of all states with
	respect to $r$. The excitation range of the Rydberg laser is represented by
	the position and width ($\sim1$~MHz) of the red line. (b) States in (a)
	with line darkness weighted by the oscillator strength to the ground state
	(6$S_{1/2}$, $m_F = 0$) for linearly polarized light on the y-axis
	(\autoref{fig:experimentSetup}).  We choose to operate at
	$R = \unit{6.6(3)}{\micro\meter}$ to obtain a $6.4$-MHz blockade shift.
	The calculation includes a background electric field, $|\efield| = 1.6$~V/m,
	and a bias magnetic field, $B_x = 4.8$~G as is used in our experiment.
	}
	\label{fig:blockadeCalc}
\end{figure}

We observe variation in surface charging with changes in background cesium vapor
pressure and dipole trap laser intensity. While the dipole trap laser drives
charge production, whose steady-state value increases with laser intensity, the
cesium vapor pressure modifies properties of the charging process.  We observe
that increasing vapor pressure decreases the charging time constant, which
ranges from minutes to hours, and additionally reduces the field strength.  It is
known that the density of cesium coverage on a surface modifies the work
function \cite{Taylor1933,Chou2012_JOPCM}. Changes to the work function would
affect the laser-induced charging and is a likely explanation for the observed
trend with cesium vapor pressure.
To reduce fluctuations in $\efield$, we stabilize the dipole trap laser power
and cesium vapor pressure. Even so, this does not eliminate the observed
fluctuations in Rydberg state energy. The calculated electric field perturbation
approximately follows the quadratic Stark effect as is shown in
\autoref{fig:charging}. Noise in the Stark shift, $\delta E_{\text{Stark}}$,
increases linearly with electric field noise, $\delta\efield$, or
\begin{equation*}
	\delta E_{\text{Stark}} = 2\alpha_{r} |\efield|\delta\efield.
\end{equation*}
Therefore, to reduce resonance fluctuations, we introduce a charging laser beam
to the experiment to gain leverage over the electric field environment.

The charging beam generates charge on the ITO glass plates at a position chosen
to counter act the electric field at the atom as shown in
\autoref{fig:experimentSetup}. The exact position and intensity of the charging
beam is finely tuned to minimize the electric field at the atom. The result of
this process is found in \autoref{fig:charging}(b), where the spectrum shifts
blue and the Stark splitting is reduced. From the measured Stark splitting, we
estimate that the electric field at the atom is 1.5(1)~V/m. With the
introduction of the charging beam, the measured Rydberg resonance has a
full width at half maximum of 440(50)~kHz. Additionally, measurements of the
spectrum over a 9-h time period indicate a characteristic resonance drift of
20~kHz over $30$~min.  We find that the combination of the ITO coated
surfaces enables this stability in our system for high principal quantum number
and correspondingly high electric field sensitivity.  This is a 100-fold
improvement when compared with our previous studies that utilized an identical
lens with AR coating on the surface closest to the atom. This improved stability
allows for coherent excitation of blockaded Rydberg atoms as is demonstrated in
\autoref{sec:rydbergBlockade}.

\section{Rydberg Blockade} \label{sec:rydbergBlockade}

The Rydberg blockade effect occurs when the EDDI potential energy
$U_{\text{int}}$ between nearby atoms shifts the doubly excited Rydberg state
$|rr\rangle$ out of resonance with the excitation laser, blocking multiple
Rydberg excitations. Producing efficient blockade entails maximizing the
interaction potential energy so that $U_{\text{int}} \gg \Omega$. At large
interatomic separation $R$ the interaction obeys a van der Waals potential of the
form $U_{\text{int}} = C_6/R^6$. From this we can increase $U_{\text{int}}$ by
either increasing principal quantum number since $C_6 \propto n^{11}$
\cite{SaffmanWalker2005_PRA} or decreasing $R$. While it is attractive to
maximize $U_{\text{int}}$ by increasing $n$, the electric polarizability also
rises as $n^{7}$ causing the system to be more susceptible to stray electric
fields.  We consequently target a value of $n$ with manageable dc Stark shift
and significant blockade at interatomic distances that are optically resolvable.
While we have already shown that the former condition is satisfied for $n=84$ in
\autoref{sec:efield}, the latter can be determined numerically. Therefore, to
select a Rydberg state and interatomic separation that satisfies $U_{\text{int}}
\gg \Omega$, we numerically calculate the doubly excited Rydberg state spectrum
$nP+nP$ as a function of interatomic separation.

The numerical calculation determines the energies and transition oscillator
strengths of the sublevels contained within $nP+nP$ as functions of
$R$. The calculation diagonalizes a Hamiltonian that contains electric dipole
and quadrupole interactions in a chosen subspace of $nl+n'l'$ state pairs where
the primed variables refer to the state of the second atom. As an example, we
perform the calculation for the 84$P_{3/2}$ states.  Because interaction
channels $nl+n'l'$ with energies closest to $84P_{3/2} + 84P_{3/2}$
are the largest contributors to the EDDI \cite{SaffmanWalker2010}, we perform
the calculation for states that span a 40-GHz range centered around this target
state and restrict $l$ and $l'$ to 0--5.  We find the inclusion of the high
angular momentum states necessary as the spectrum becomes heavily mixed and
therefore choose the largest range allowed for by our current computational
resources. Techniques for computing $U_{\text{int}}$ are found in
\cite{SaffmanWalker2005_PRA,SaffmanWalker2008,Shaffer2006_PRA}, and the result
closely resembles \cite{Shaffer2006_PRA}. We calculate that blockade becomes
significant below \unit{7}{\micro\meter} for $\Omega/2\pi = 1$~MHz as is shown
in \autoref{fig:blockadeCalc}.  For the experiment we choose to use a mean
separation of \unit{6.6(3)}{\micro\meter} resulting in $U_{\text{int}}/2\pi\sim
6.4$~MHz. The calculation of nonzero blockade for all sublevels appears to be
in contradiction with the prediction of F\"orster zeros (unshifted states)
\cite{SaffmanWalker2005_JPhysB} for the $n P_{3/2}+nP_{3/2}$ state
studied here. The difference arises because we are concerned with values of $R$
where convergence of the calculation requires multiple $nl + nl\rightarrow n'l'
+ n''l''$ channels, whereas \cite{SaffmanWalker2005_JPhysB} emphasizes the most
significant term at large interatomic separations (van der Waals regime). With
an interaction potential on the order of 1~MHz, signatures of Rydberg blockade
should be present.

\begin{figure}%[htbp]
	\includegraphics{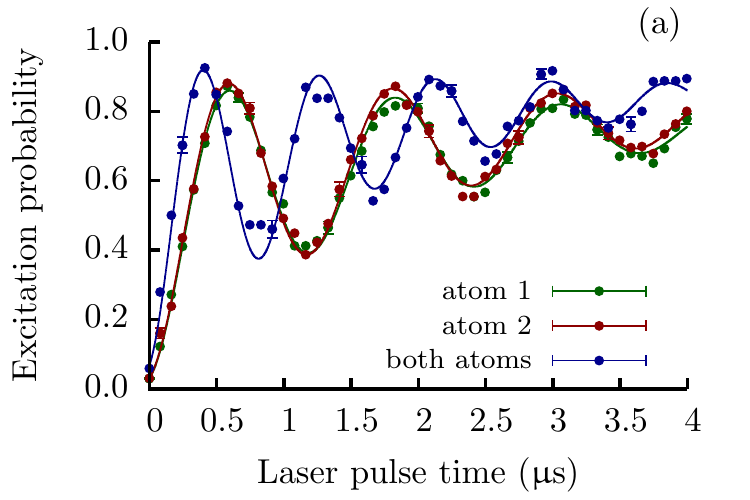}
	
	\includegraphics{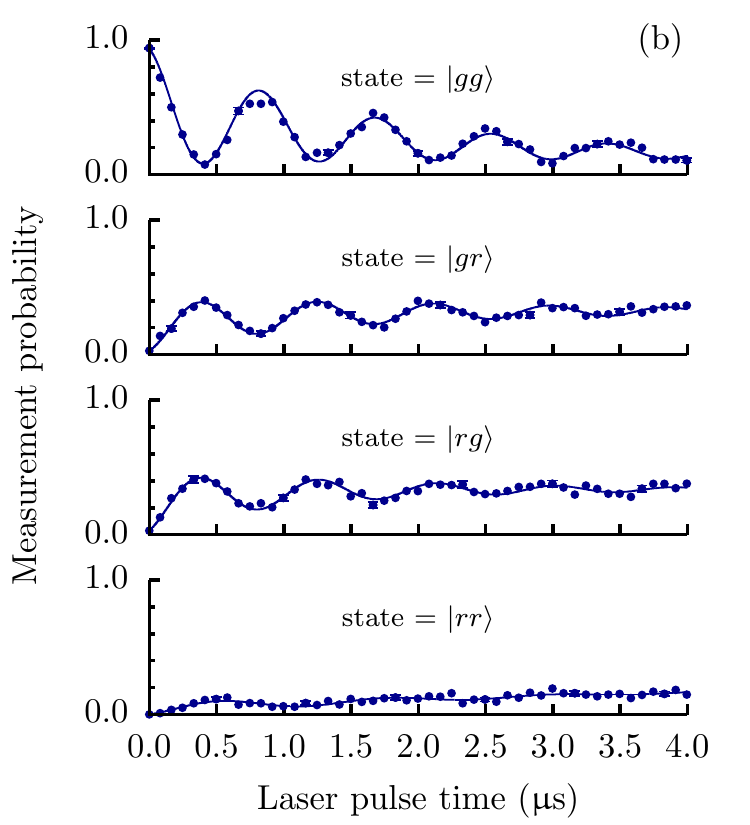}
	\caption{ (Color online)
	Observation of coherent excitation and Rydberg blockade. (a) Excitation
	probability for single, noninteracting atoms (atoms 1 and 2) and two
	interacting atoms (both atoms). The noninteracting data are labeled atom 1
	and atom 2, where the number represents exclusive loading of either the
	first or second dipole trap. Here, the excitation probability $P_{\text{e}}$
	is defined as $P_{\text{e}} = 1-P(|g\rangle)$ for single atoms and
	$P_{\text{e}} = 1 - P(|gg\rangle)$ for the two-atom case. This plot
	highlights the $\sqrt{2}$ increase in excitation Rabi frequency of two atoms
	in the strong blockade regime. (b) Measured evolution of the blockaded
	two-atom system in the basis $\{|gg\rangle, |gr\rangle, |rg\rangle,
	|rr\rangle\}$. Error bars shown are representative.
	}
	\label{fig:rydbergBlockade}
\end{figure}

Using our calculation as a guide, we experimentally identify signatures of
Rydberg blockade. For the experiment, we resonantly excite $|84P_{3/2}$,
$m_j = 3/2\rangle$ with variable laser pulse duration. We use a 319-nm laser
intensity of 9.3~kW/cm$^2$ to achieve $\Omega = 0.816(4)$~MHz and apply a 4.8-G
bias magnetic field to break the degeneracy in $m_j$.  The measured Rydberg
excitation probability with single as well as two interacting atoms is shown in
\autoref{fig:rydbergBlockade}(a). The measured ratio of the Rydberg excitation
Rabi frequencies $\Omega_{\text{2-atoms}}/\Omega$ is 1.42(2), which is consistent
with $\sqrt{2}$. An increase in the excitation Rabi frequency of $\sqrt{2}$ is
expected in the strongly blockaded regime where the system oscillates between
the ground state and a state that collectively shares a single Rydberg
excitation, as was first observed by Gaetan et
al.~\cite{Grangier2009_NaturePhys}. Additional evidence of Rydberg blockade is
shown in \autoref{fig:rydbergBlockade}(b) where we plot the two-atom evolution
in the basis $\{|gg\rangle,|gr\rangle,|rg\rangle,|rr\rangle\}$. Here, we show
that population transfer between the ground state and the singly excited state
dominates the system evolution, whereas excitation to $|rr\rangle$ is strongly
suppressed. Both plots illustrate coherent control of two strongly blockaded
atoms.

The coherent dynamics of this system, with or without two-atom blockade,
indicates decoherence dominated by population relaxation $\Gamma_{\text{loss}}$
out of the Rydberg state.  We measure $\Gamma_{\text{loss}} = 1.2(1)$~MHz, which
is substantially broader than the calculated state linewidth of
4~kHz~\footnote{Calculation includes the effect of room temperature blackbody
radiation.}. The trend in the evolution of the atom towards excitation to
$|r\rangle$, shown in \autoref{fig:rydbergBlockade}, occurs because our
detection method can not differentiate between population in $|r\rangle$ and
other atom-loss mechanisms. Examples of possible loss sources include, an
applied force on the center of mass of the atom from an electric field gradient
and a reduced state lifetime due to background RF fields.  Order-of-magnitude
estimates suggest that the latter example is more likely. In future work, we aim
to investigate the source of $\Gamma_{\text{loss}}$ and mitigate its decohering
effects.

\section{Summary and Outlook}
\label{sec:conclusion}

In summary, we present experiments and theoretical models that focus on Rydberg
blockade with a single-photon transition. We construct a UV laser for direct
excitation to $n\text P_{3/2}$ Rydberg states and demonstrate the accuracy of
our calculations for the Rydberg spectrum and oscillator strength with
single-atom spectroscopy. These Rydberg atoms are employed as electric field
sensors to study laser-induced charging of nearby surfaces, and we utilize this
information to mitigate noise on the Rydberg resonance frequency due to the dc
Stark effect.  Finally, we model EDDIs for two $84\text P_{3/2}$ Rydberg atoms
as a function of interatomic separation and demonstrate Rydberg blockade through
an increase in the collective excitation Rabi frequency.

In principle, this single-photon approach offers an advantage over two-photon
Rydberg excitation by eliminating the need for an intermediate state, thus
avoiding channels for photon scattering, frequency noise, and dipole forces.
Reducing photon scattering is especially attractive for the study of dipolar
interactions between Rydberg dressed states, where allowing the system to relax
to equilibrium can take on order of 1~ms \cite{Rolston2010PRA,Keating2013PRA}.
However, this fundamental limit on the photon scattering rate remains to be
demonstrated in our experiment. The moderate increase in Doppler sensitivity
when using single-photon Rydberg excitation can be addressed with more advanced
cooling techniques, such as ground-state cooling
\cite{Regal2012_PRX,Lukin2013_PRL}.

\section{Acknowledgements}
We thank Mark Saffman, Antoine Browaeys, James Shaffer, Steve Rolston, Ivan
Deutsch, Tyler Keating, and Rob Cook for helpful discussions and suggestions. We
would also like to thank George Burns, Peter Schwindt, Michael Mangan, Cort
Johnson, and Andrew Ferdinand for contributions to the experiment. We
acknowledge Laboratory Directed Research and Development for funding this work.
Sandia National Laboratories is a multi-program laboratory managed and operated
by Sandia Corporation, a wholly owned subsidiary of Lockheed Martin Corporation,
for the U.S.\ Department of Energy's National Nuclear Security Administration
under Contract No.\ DE-AC04-94AL85000.

\bibliographystyle{apsrev4-1}
\bibliography{direct_excitation}

\end{document}